\def\be{\begin{equation}}
\def\ee{\end{equation}}
\def\ba{\begin{array}}

\def\ea{\end{array}}

\documentclass[12pt]{article}
\usepackage{amsfonts,amssymb,amsmath,graphicx}
\newtheorem{theorem}{Theorem}

\begin{document}
\parskip=3pt
\parindent=18pt
\baselineskip=20pt
\setcounter{page}{1}
\centerline{\large\bf Estimation on geometric measure of quantum coherence}
\vspace{6ex}
\centerline{{\sf Hai-Jun Zhang,$^{\natural,\S}$}
\footnote{\sf zhnavy77@126.com}
~~~ {\sf Bin Chen,$^{\star,\dag}$}
\footnote{\sf chenbin5134@163.com}
~~~ {\sf Ming Li,$^{\S}$}
\footnote{\sf liming3737@163.com}
~~~ {\sf Shao-Ming Fei,$^{\natural,\sharp}$}
\footnote{\sf feishm@cnu.edu.cn}
~~~ {\sf Gui-Lu Long$^{\star,\dag,\ddag}$}
\footnote{\sf gllong@tsinghua.edu.cn}
}
\vspace{4ex}
\centerline
{\it $^\natural$ School of Mathematical Sciences, Capital Normal University, Beijing 100048, China}\par
\centerline
{\it $^\S$ College of the Science, China University of Petroleum, Qingdao, 266580, China}\par
\centerline
{\it $^\star$ State Key Laboratory of Low-Dimensional Quantum Physics and Department of Physics,}\par
\centerline
{\it Tsinghua University, Beijing 100084, China}\par
\centerline
{\it $^\dag$ Tsinghua National Laboratory for Information Science and Technology, Beijing 100084, China}\par

\centerline
{\it $^\sharp$ Max-Planck-Institute for Mathematics in the Sciences, 04103 Leipzig, Germany}\par
\centerline
{\it $^\ddag$ Collaborative Innovation Center of Quantum Matter, Beijing 100084, China}\par
\vspace{6.5ex}
\parindent=18pt
\parskip=5pt
\begin{center}
\begin{minipage}{5in}
\vspace{3ex}
\centerline{\large Abstract}
\vspace{4ex}
We study the geometric measure of quantum coherence recently proposed in [Phys. Rev. Lett. 115, 020403 (2015)].
Both lower and upper bounds of this measure are provided.
These bounds are shown to be tight for a class of important coherent states -- maximally coherent mixed states. The trade-off relation between quantum coherence and mixedness for this measure
is also discussed.

\end{minipage}
\end{center}

\newpage

\section{Introduction}

Quantum coherence plays a vital role in quantum physics and quantum information processing.
Being an important physical resource, it is tightly related to various research fields such as low-temperature thermodynamics \cite{y1,y2,y3,y4,y5}, quantum biology \cite{a1,y6,y7,y8,y9,y10,y11}, nanoscale physics \cite{y12,y13}, ect.
Formulating the theory of quantum coherence is a long-standing problem and considerable progress have been made in quantum optics \cite{y14,y15,a2}.
Recently, a rigorous framework for the quantification of coherence was introduced \cite{Baumgratz}.
Inspired by the quantitative theory of quantum entanglement, the authors in Ref. \cite{Baumgratz} provided a framework to quantify coherence by defining the so-called incoherent states and incoherent operations.
These two concepts are analogous to the separable states and local operations and classical communication (LOCC) repectively in the quantum entanglement theory.

Fixing a basis $\{|i\rangle\}_{i=1}^{d}$ of a $d$-dimensional Hilbert space $\mathcal{H}$, the incoherent states are defined as \cite{Baumgratz}:
\begin{equation}\label{e1}
\sigma=\sum_{i=1}^{d}p_{i}|i\rangle\langle i|,
\end{equation}
where $p_{i}\geq0,\sum_{i=1}^{d}p_{i}=1$.
Quantum states that cannot be written in the above form are called coherent states.
Let $\Lambda$ be a completely positive trace preserving (CPTP) map,
\begin{equation}\label{e2}
\Lambda(\rho)=\sum_{n}K_{n}\rho K_{n}^{\dag},
\end{equation}
where $\{K_{n}\}$ is a set of Kraus operators satisfying $\sum_{n}K_{n}^{\dag}K_{n}=\mathbb{I}_{d}$.
Let $\mathcal{I}$ be the set of incoherent states.
If $K_{n}\mathcal{I}K_{n}^{\dag}\subseteq\mathcal{I}$ for all $n$, then $\{K_{n}\}$ is called a set of incoherent Kraus operators, and the corresponding $\Lambda$ is called an incoherent operation \cite{Baumgratz}.
Obviously, $\Lambda(\mathcal{I})\subseteq\mathcal{I}$.
Similar to the quantification of quantum entanglement \cite{qe1,qe2,qe3,qe4}, Baumgratz \emph{et al.} proposed the following conditions to be satisfied as a measure of coherence $C(\rho)$ \cite{Baumgratz}:

(1) $C(\rho)\geq0$, and $C(\rho)=0$ if and only if $\rho\in\mathcal{I}$;

(2) $C(\Lambda(\rho))\leq C(\rho)$ for any incoherent operation $\Lambda$;

(3) $\sum_{n}p_{n}C(\rho_{n})\leq C(\rho)$, where $p_{n}=\mathrm{Tr}(K_{n}\rho K_{n}^{\dag}),\rho_{n}=K_{n}\rho K_{n}^{\dag}/p_{n},\{K_{n}\}$ is a set of incoherent Kraus operators;

(4) $C(\sum_{i}p_{i}\rho_{i})\leq\sum_{i}p_{i}C(\rho_{i})$ for any set of quantum states $\{\rho_{i}\},p_{i}\geq0,\sum_{i=1}^{d}p_{i}=1$.
It is obvious that conditions (3) and (4) imply condition (2).
A quantity that satisfies conditions (1), (2), and (3) is called a coherence monotone.
If it satisfies condition (4) in addition, then we call it a convex coherence monotone.

Distance-based measures are the best options for quantifying coherence, which are defined as $C_{\mathcal{D}}(\rho)=\min_{\sigma\in\mathcal{I}}\mathcal{D}(\rho,\sigma)$.
If $\mathcal{D}$ is the $l_{1}$-norm \cite{Baumgratz}, then $C_{l_{1}}(\rho)=\sum_{i\neq j}|\rho_{ij}|$, where $\rho_{ij}=\langle i|\rho|j\rangle$.
If $\mathcal{D}$ is the quantum relative entropy \cite{Baumgratz}, i.e., $\mathcal{D}(\rho,\sigma)=S(\rho\|\sigma)=\mathrm{Tr}(\rho\ln\rho)-\mathrm{Tr}(\rho\ln\sigma)$, then $C_{rel}(\rho)=S(\rho_{d})-S(\rho)$,
where $\rho_{d}=\sum_{i=1}^{d}\rho_{ii}|i\rangle\langle i|$, and $S(\rho)=-\mathrm{Tr}(\rho\ln\rho)$ is the von Neumann entropy.
For other distance-based measures, many results have been obtained so far \cite{Girolami,Shao,Rana,Streltsov}.

In Ref. \cite{Streltsov}, the authors provided an operational link between coherence and entanglement.
They showed that for any (convex) entanglement monotone $E$, one can define a corresponding (convex) coherence monotone $C_{E}$ via an explicit formula.
As an example, they proved that the geometric measure of coherence $C_{g}$, defined by the fidelity-based geometric measure of entanglement \cite{SKB},
is indeed a convex coherence monotone.
The expression of $C_{g}$ has been derived as $C_{g}(\rho)=1-\max_{\sigma\in\mathcal{I}}F(\rho,\sigma)$, where $F(\rho,\sigma)=(\mathrm{Tr}\sqrt{\sqrt{\sigma}\rho\sqrt{\sigma}})^{2}$ is the fidelity of two density operators $\rho$ and $\sigma$,
and analytical formula of $C_{g}(\rho)$ for any single-qubit state $\rho$ is given \cite{Streltsov}.
However, for an arbitrary qudit state, computation of $C_{g}(\rho)$ is formidably difficult. Just like estimation of concurrence and entanglement of formation, it is also important to estimate the lower and upper bounds of $C_{g}(\rho)$.

Recently, Singh \emph{et al.} studied quantum coherence and mixedness in any $d$-dimensional quantum system \cite{Singh}.
They have shown that for a fixed mixedness, the amount of coherence is restricted.
A trade-off relation between coherence quantified by the $l_{1}$ norm and mixedness quantified by the normalized linear entropy is provided \cite{Singh}.
This result gives rise to the maximally coherent mixed states (MCMS), i.e., quantum states with maximal coherence and mixedness.
They also discussed such trade-off relation and the form of MCMS for any qubit systems when the geometric coherence and geometric mixedness are considered.
However, for any qudit system, the form of MCMS for geometric coherence is still unknown.

In this paper, we derive lower and upper bounds for the geometric measure of coherence for arbitrary dimension $d$, by using the concepts of sub-fidelity and super-fidelity introduced in Ref. \cite{Miszczak,Chen,Wu1,Wu2,Wu3}.
These bounds are shown to be tight for a class of maximally coherent mixed states.
Furthermore, we also discuss the form of MCMS for the geometric measure of coherence for any qudit systems.

\section{Estimation of geometric measure of coherence}
Let $\{|i\rangle\}_{i=1}^{d}$ be a fixed basis of a $d$-dimensional Hilbert space.
The incoherent states are represented as $\sigma=\sum_{i=1}^{d}x_{i}|i\rangle\langle i|$, where $x_{i}\geq0,\sum_{i=1}^{d}x_{i}=1$.
Any density operator $\rho$ is written as $\rho=\sum_{i,j=1}^{d}\rho_{ij}|i\rangle\langle j|$.

In Ref. \cite{Miszczak}, the authors introduced two quantities $E(\rho,\sigma)$ and $G(\rho,\sigma)$ called sub-fidelity and super-fidelity, respectively, as the lower and upper bounds of the fidelity $F(\rho,\sigma)$ for two quantum states $\rho$ and $\sigma$. They are defined as
\begin{equation}
\begin{split}
E(\rho,\sigma)&=\mathrm{Tr}(\rho\sigma)+\sqrt{2[(\mathrm{Tr}(\rho\sigma))^{2}-\mathrm{Tr}(\rho\sigma\rho\sigma)]},\\
G(\rho,\sigma)&=\mathrm{Tr}(\rho\sigma)+\sqrt{1-\mathrm{Tr}(\rho^{2})}\sqrt{1-\mathrm{Tr}(\sigma^{2})}.
\end{split}
\end{equation}
It holds that $E(\rho,\sigma)\leq F(\rho,\sigma)\leq G(\rho,\sigma)$,
and all three quantities are equal when $d=2$, or at least one of $\rho$ and $\sigma$ is a pure state.
Sub- and super-fidelity have many elegant properties, such as bounded, i.e., $0\leq E(\rho,\sigma)\leq1,0\leq G(\rho,\sigma)\leq1$, symmetry, unitary invariance, concavity, ect.
Here we use them to estimate the geometric measure of coherence $C_{g}(\rho)=1-\max_{\sigma\in\mathcal{I}}F(\rho,\sigma)$.
It is obvious that $1-\max_{\sigma\in\mathcal{I}}G(\rho,\sigma)\leq C_{g}(\rho)\leq1-\max_{\sigma\in\mathcal{I}}E(\rho,\sigma)$.
We only need to find the maximal value of $E(\rho,\sigma)$ and $G(\rho,\sigma)$, when $\sigma$ run over all the incoherent states.
We have the following Theorem.
\begin{theorem}\label{t1}
For any density operator $\rho$ acting on a $d$-dimensional Hilbert space, the geometric measure of coherence $C_{g}(\rho)$ satisfies the following inequality:
\begin{equation}\label{et1}
1-\frac{1}{d}-\frac{d-1}{d}\sqrt{1-\frac{d}{d-1}\left(\mathrm{Tr}(\rho^{2})-\sum_{i=1}^{d}\rho_{ii}^{2}\right)}\leq C_{g}(\rho)\leq1-\max_{i}\{\rho_{ii}\},
\end{equation}
and $C_{g}(\rho)=1-\max_{i}\{\rho_{ii}\}$ when $\rho$ is a pure state.
\end{theorem}

\emph{Proof.} If $\rho$ is a pure state, then $F(\rho,\sigma)=\mathrm{Tr}(\rho\sigma)=\sum_{i=1}^{d}\rho_{ii}x_{i}\leq\max_{i}\{\rho_{ii}\}$.
Thus we have $C_{g}(\rho)=1-\max_{i}\{\rho_{ii}\}$.
Next, we suppose that $\rho$ is a mixed state.

We first estimate $\max_{\sigma\in\mathcal{I}}E(\rho,\sigma)$.
Noting that $\mathrm{Tr}(\rho\sigma\rho\sigma)=\sum_{i,j=1}^{d}|\rho_{ij}|^{2}x_{i}x_{j}$,
we have
\begin{equation}\label{ep1}
\begin{split}
E(\rho,\sigma)&=\sum_{i=1}^{d}\rho_{ii}x_{i}+\sqrt{2}\left[\sum_{i,j=1}^{d}\left(\rho_{ii}\rho_{jj}-|\rho_{ij}|^{2}\right)x_{i}x_{j}\right]^{\frac{1}{2}}\\
&:=f(x_{1},\ldots,x_{d}).
\end{split}
\end{equation}
To find the maximal value of $f(x_{1},\ldots,x_{d})$ over the closed domain $\overline{D}=\{(x_{1},\ldots,x_{d})\in\mathbb{R}^{d}:x_{i}\geq0,\sum_{i=1}^{d}x_{i}=1\}$,
we first assume that the maximal value is achieved in the open domain $D=\{(x_{1},\ldots,x_{d})\in\mathbb{R}^{d}:x_{i}>0,\sum_{i=1}^{d}x_{i}=1\}$,
then the maximum point satisfies
\begin{equation}\label{ep2}
\frac{\partial f}{\partial x_{i}}=\rho_{ii}+\frac{\sqrt{2}\sum_{j=1}^{d}\left(\rho_{ii}\rho_{jj}-|\rho_{ij}|^{2}\right)x_{j}}
{\sqrt{\sum_{i,j=1}^{d}\left(\rho_{ii}\rho_{jj}-|\rho_{ij}|^{2}\right)x_{i}x_{j}}}=0,~~~\forall i.
\end{equation}
From $\sum_{i=1}^{d}\frac{\partial f}{\partial x_{i}}x_{i}=0$, we get
$\sum_{i=1}^{d}\rho_{ii}x_{i}+\sqrt{2}\left[\sum_{i,j=1}^{d}\left(\rho_{ii}\rho_{jj}-|\rho_{ij}|^{2}\right)x_{i}x_{j}\right]^{\frac{1}{2}}=0$.
That is to say, the maximal value of $f$ is equal to zero, which is impossible.
Thus we can confirm that the maximum point of $f$ must be on the boundary of $\overline{D}$, and $f_{\max}$ is no less than $f(\textbf{x})$ when
$\textbf{x}\in D_{1}\subseteq\overline{D}$, where
$D_{1}=\{(x_{1},\ldots,x_{d})\in\mathbb{R}^{d}:\exists ~k, ~\mathrm{s.t.} ~x_{k}=1, ~\mathrm{and} ~x_{i}=0, \forall i\neq k\}$.
Therefore we have $f_{\max}\geq\max_{i}\{\rho_{ii}\}$, and $1-\max_{\sigma\in\mathcal{I}}E(\rho,\sigma)\leq1-\max_{i}\{\rho_{ii}\}$.

We now compute $\max_{\sigma\in\mathcal{I}}G(\rho,\sigma)$.
Note that
\begin{equation}\label{ep3}
\begin{split}
G(\rho,\sigma)&=\sum_{i=1}^{d}\rho_{ii}x_{i}+\sqrt{1-\mathrm{Tr}(\rho^{2})}\sqrt{1-\sum_{i=1}^{d}x_{i}^{2}}\\
&:=g(x_{1},\ldots,x_{d}).
\end{split}
\end{equation}
To find the maximal value of $g$ over $\overline{D}$, we use the Lagrange multiplier method.
Let $L(x_{1},\ldots,x_{d})=\sum_{i=1}^{d}\rho_{ii}x_{i}+\sqrt{1-\mathrm{Tr}(\rho^{2})}\sqrt{1-\sum_{i=1}^{d}x_{i}^{2}}+\lambda(\sum_{i=1}^{d}x_{i}-1)$.
Then we have
\begin{equation}\label{ep4}
\frac{\partial L}{\partial x_{i}}=\rho_{ii}-\sqrt{1-\mathrm{Tr}(\rho^{2})}\frac{x_{i}}{\sqrt{1-\sum_{i=1}^{d}x_{i}^{2}}}+\lambda=0,~~~\forall i.
\end{equation}
This implies that
\begin{equation}\label{ep5}
(\rho_{ii}+\lambda)\sqrt{1-\sum_{i=1}^{d}x_{i}^{2}}=x_{i}\sqrt{1-\mathrm{Tr}(\rho^{2})},~~~\forall i.
\end{equation}
Summing over the above equations from $1$ to $d$, we get
\begin{equation}\label{ep6}
(1+d\lambda)\sqrt{1-\sum_{i=1}^{d}x_{i}^{2}}=\sqrt{1-\mathrm{Tr}(\rho^{2})}.
\end{equation}
Thus
\begin{equation}\label{ep7}
\rho_{ii}+\lambda=(1+d\lambda)x_{i},~~~\forall i.
\end{equation}
From Eq. (\ref{ep6}) and Eq. (\ref{ep7}), we obtain
\begin{equation}\label{ep8}
\sum_{i=1}^{d}x_{i}^{2}=1-\frac{1-\mathrm{Tr}(\rho^{2})}{(1+d\lambda)^{2}}=\sum_{i=1}^{d}\left(\frac{\rho_{ii}+\lambda}{1+d\lambda}\right)^{2},
\end{equation}
which immediately yields that
\begin{equation}\label{ep9}
\lambda=-\frac{1}{d}\pm\frac{1}{d}\sqrt{1-\frac{d}{d-1}\left(\mathrm{Tr}(\rho^{2})-\sum_{i=1}^{d}\rho_{ii}^{2}\right)}.
\end{equation}
We can also derive from Eq. (\ref{ep6}) and Eq. (\ref{ep7}) that
\begin{equation}
\begin{split}
g(x_{1},\ldots,x_{d})&=\sum_{i=1}^{d}\left[(1+d\lambda)x_{i}-\lambda\right]x_{i}+\frac{1-\mathrm{Tr}(\rho^{2})}{1+d\lambda}\\
&=1+(d-1)\lambda.
\end{split}
\end{equation}
Hence the maximal value of $g$ over $\overline{D}$ is equal to $\frac{1}{d}+\frac{d-1}{d}\sqrt{1-\frac{d}{d-1}\left(\mathrm{Tr}(\rho^{2})-\sum_{i=1}^{d}\rho_{ii}^{2}\right)}$ by Eq. (\ref{ep9}).
Therefore we have
$1-\max_{\sigma\in\mathcal{I}}G(\rho,\sigma)=1-\frac{1}{d}-\frac{d-1}{d}\sqrt{1-\frac{d}{d-1}\left(\mathrm{Tr}(\rho^{2})-\sum_{i=1}^{d}\rho_{ii}^{2}\right)}$
for any mixed state.
This completes the proof.  \quad $\Box$

For $d=2$, if $\rho$ is a pure state, then $C_{g}(\rho)=1-\max\{\rho_{11},\rho_{22}\}$.
Taking into account that $\mathrm{Tr}(\rho^{2})=1$, we have $C_{g}(\rho)=\frac{1}{2}-\frac{1}{2}\sqrt{1-4|\rho_{12}|^{2}}$.
If $\rho$ is a mixed state, then we have $C_{g}(\rho)=1-\max_{\sigma\in\mathcal{I}}G(\rho,\sigma)=\frac{1}{2}-\frac{1}{2}\sqrt{1-4|\rho_{12}|^{2}}$.
Therefore $C_{g}(\rho)=\frac{1}{2}-\frac{1}{2}\sqrt{1-4|\rho_{12}|^{2}}$ holds for any qubit state $\rho$, which coincides with the result obtained in Ref. \cite{Streltsov}.

As an example of Theorem \ref{t1}, let us consider a class of coherent states -- maximally coherent mixed states (MCMS) \cite{Singh}, which are defined as
\begin{equation}\label{c0}
\rho_{m}=p|\psi_{d}\rangle\langle\psi_{d}|+\frac{1-p}{d}\mathbb{I}_{d},
\end{equation}
where $0<p\leq1$, and $|\psi_{d}\rangle=\frac{1}{\sqrt{d}}\sum_{i=1}^{d}|i\rangle$ is the maximally coherent state.
From (\ref{et1}), we get
\begin{equation}\label{c1}
1-\frac{1}{d}-\frac{d-1}{d}\sqrt{1-p^{2}}\leq C_{g}(\rho_{m})\leq1-\frac{1}{d}.
\end{equation}

We now compute $C_{g}(\rho_{m})$.
Note that
\begin{equation}\label{c2}
\sqrt{\sigma}\rho_{m}\sqrt{\sigma}=\frac{1}{d}\sum_{i}x_{i}|i\rangle\langle i|+\frac{p}{d}\sum_{i\neq j}\sqrt{x_{i}x_{j}}|i\rangle\langle j|,
\end{equation}
and suppose that
\begin{equation}\label{c3}
\sqrt{\sqrt{\sigma}\rho_{m}\sqrt{\sigma}}=a\sum_{i}\sqrt{x_{i}}|i\rangle\langle i|+b\sum_{i,j}\sqrt{x_{i}x_{j}}|i\rangle\langle j|,~~~a\geq0,b\geq0.
\end{equation}
Then we have
\begin{equation}
\begin{split}
\frac{1}{d}\sum_{i}x_{i}|i\rangle\langle i|+\frac{p}{d}\sum_{i\neq j}\sqrt{x_{i}x_{j}}|i\rangle\langle j|&=\sum_{i}\left[\left(a^{2}+b^{2}\right)x_{i}+2abx_{i}\sqrt{x_{i}}\right]|i\rangle\langle i|\\
&+\sum_{i\neq j}\left[b^{2}\sqrt{x_{i}x_{j}}+ab\left(x_{i}\sqrt{x_{j}}+x_{j}\sqrt{x_{i}}\right)\right]|i\rangle\langle j|.
\end{split}
\end{equation}
Comparing both sides of the above equation, we obtain
\begin{equation}\label{c4}
\begin{split}
\left(a^{2}+b^{2}\right)+2ab\sqrt{x_{i}}&=\frac{1}{d},~~~\forall i,\\
b^{2}+ab\left(\sqrt{x_{j}}+\sqrt{x_{i}}\right)&=\frac{p}{d},~~~i\neq j.
\end{split}
\end{equation}
Summing over the above equations from 1 to $d$, respectively, we have
\begin{equation}
\begin{split}
d\left(a^{2}+b^{2}\right)+2ab\sum_{i}\sqrt{x_{i}}&=1,\\
db^{2}+2ab\sum_{i}\sqrt{x_{i}}&=p,
\end{split}
\end{equation}
which yields that $a=\sqrt{\frac{1-p}{d}}$.
On the other hand, it holds that $\sum_{i=1}^{d}\left[\frac{1}{d}-\left(a^{2}+b^{2}\right)\right]^{2}=4a^{2}b^{2}\sum_{i=1}^{d}x_{i}=4a^{2}b^{2}$ from Eq. (\ref{c4}), then we get $b=\frac{1}{d}\left(\sqrt{1-p+dp}-\sqrt{1-p}\right)$.
Thus we have
\begin{equation}
\begin{split}
\left(\mathrm{Tr}\sqrt{\sqrt{\sigma}\rho_{m}\sqrt{\sigma}}\right)^{2}&=\left(a\sum_{i}\sqrt{x_{i}}+b\right)^{2}\\
&\leq\left(a\sqrt{d}+b\right)^{2}\\
&=\left[\sqrt{1-p}+\frac{1}{d}\left(\sqrt{1-p+dp}-\sqrt{1-p}\right)\right]^{2}
\end{split}
\end{equation}
by use of Cauchy-Schwarz inequality, and the equality holds if and only if $x_{i}=\frac{1}{d},\forall i$.
Therefore we have
\begin{equation}\label{c5}
C_{g}(\rho_{m})=1-\left[\sqrt{1-p}+\frac{1}{d}\left(\sqrt{1-p+dp}-\sqrt{1-p}\right)\right]^{2}.
\end{equation}

\begin{figure}
\centering
\includegraphics[width=7cm]{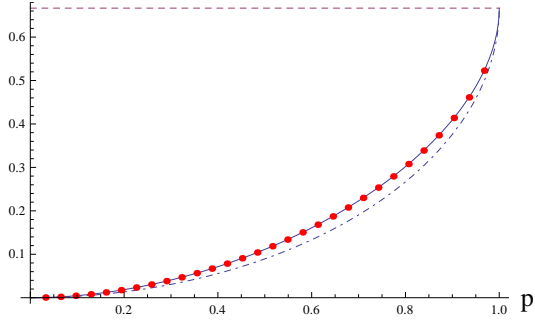}
\caption{The solid line is the value of $C_{g}(\rho_{m})$ for $0<p\leq1$.
The dot-dashed line is the lower bound in (\ref{et1}). The dashed line is the upper bound in (\ref{et1}), and the dotted line is the upper bound in (\ref{et2}). It can be seen that the upper bound in (\ref{et2}) is achieved for $\rho_{m}$, and it is much tighter than the one in (\ref{et1}) in this case.}\label{goc}
\end{figure}

For $d=3$, the comparison between $C_{g}(\rho_{m})$ and the lower and upper bounds of $C_{g}(\rho_{m})$ in (\ref{c1}) is shown in FIG \ref{goc}.

It seems that the upper bound we derived in Theorem 1 is large for mixed states.
However, we have the following improved upper bound.
\begin{theorem}\label{t2}
Let $\sqrt{\rho}=\sum_{i,j}b_{ij}|i\rangle\langle j|$. Then we have
\begin{equation}\label{et2}
C_{g}(\rho)\leq1-\left(\sum_{i}b_{ii}^{2}\right),
\end{equation}
\end{theorem}

\emph{Proof.} Since $\sqrt{F(\rho,\sigma)}=\max_{U}\mathrm{Tr}(U\sqrt{\rho}\sqrt{\sigma})\geq\mathrm{Tr}(\sqrt{\rho}\sqrt{\sigma})$, we have
$C_{g}(\rho)=1-\max_{\sigma\in\mathcal{I}}F(\rho,\sigma)\leq1-\max_{\sigma\in\mathcal{I}}\left[\mathrm{Tr}(\sqrt{\rho}\sqrt{\sigma})\right]^{2}$.
Note that
\begin{equation}
\begin{split}
\mathrm{Tr}(\sqrt{\rho}\sqrt{\sigma})&=\sum_{i}b_{ii}\sqrt{x_{i}}\\
&\leq\sqrt{\left(\sum_{i}b_{ii}^{2}\right)\left(\sum_{i}x_{i}\right)}\\
&=\sqrt{\left(\sum_{i}b_{ii}^{2}\right)}
\end{split}
\end{equation}
by Cauchy-Schwarz inequality, and the equality holds if and only if $x_{i}=b_{ii}^{2}/(\sum_{j}b_{jj}^{2}),\forall i$.
Then we complete the proof.  \quad $\Box$

We consider again the state $\rho_{m}$ given by Eq. (\ref{c0}).
Setting $x_{i}=\frac{1}{d}, \forall i$, in Eq. (\ref{c3}), we get
\begin{equation}\label{sqm}
\sqrt{\rho_{m}}=\frac{1}{\sqrt{d}}\left(\sqrt{1-p+dp}-\sqrt{1-p}\right)|\psi_{d}\rangle\langle\psi_{d}|+\sqrt{\frac{1-p}{d}}\mathbb{I}_{d}
\end{equation}
directly.
Hence the upper bound in (\ref{et2}) for $\rho_{m}$ is equal to
\begin{equation}
1-\left[\sqrt{1-p}+\frac{1}{d}\left(\sqrt{1-p+dp}-\sqrt{1-p}\right)\right]^{2},
\end{equation}
which is exactly the value of $C_{g}(\rho_{m})$ by (\ref{c5}).
Thus we can see from FIG \ref{goc} that the upper bound in (\ref{et2}) is tighter than the one in (\ref{et1}) for some cases.

From the above discussion, we have the following Theorem.
\begin{theorem}\label{t3}
For any quantum state $\rho$, we have
\begin{equation}\label{et3}
1-\frac{1}{d}-\frac{d-1}{d}\sqrt{1-\frac{d}{d-1}\left(\mathrm{Tr}(\rho^{2})-\sum_{i=1}^{d}\rho_{ii}^{2}\right)}\leq C_{g}(\rho)\leq\min\{l_{1},l_{2}\},
\end{equation}
where $l_{1}$ and $l_{2}$ denote the upper bounds in (\ref{et1}) and (\ref{et2}), respectively. When $\rho$ is a pure state, $C_{g}(\rho)=l_{1}\leq l_{2}$.
\end{theorem}

\emph{Proof.} We only need to prove that $l_{1}\leq l_{2}$ for pure states.
Noting that $\sqrt{\rho}=\rho$ and $b_{ii}=\rho_{ii}$ for any pure state $\rho$, we have
$\sum_{i}b_{ii}^{2}=\sum_{i}\rho_{ii}^{2}\leq\max_{i}\{\rho_{ii}\}\sum_{i}\rho_{ii}=\max_{i}\{\rho_{ii}\}$.
Thus $1-\max_{i}\{\rho_{ii}\}\leq 1-(\sum_{i}b_{ii}^{2})$.  \quad $\Box$

\section{Discussion and Conclusion}
We have investigated the geometric measure of coherence.
Both lower and upper bounds of this measure have been derived.
Our upper bound can be achieved for arbitrary pure states and a class of maximally coherent mixed states.

As a matter of fact, the amount of quantum coherence is closely related to the mixedness of a quantum state.
There exits a kind of trade-off relation between the quantum coherence and the mixedness.
Recently, Singh \emph{et al.} \cite{Singh} studied the trade-off relations between coherence and mixedness for an arbitrary $d$-dimensional quantum system.
Employing the $l_{1}$-norm of coherence $C_{l_{1}}(\rho)$ and the normalized linear entropy \cite{Peters}, given by $M_{l}(\rho)=\frac{d}{d-1}(1-\mathrm{Tr}(\rho^{2}))$ as a measure of mixedness, they obtained the following inequality:
\begin{equation}\label{cm}
\frac{C_{l_{1}}^{2}(\rho)}{(d-1)^{2}}+M_{l}(\rho)\leq1.
\end{equation}
For a fixed mixedness $M_{l}$, quantum states with maximal coherence are called maximally coherent mixed states (MCMS).
It is shown that, up to incoherent unitaries, $\rho_{m}$ defined in (\ref{c0}) is the only form of MCMS with respect to the above inequality \cite{Singh}.

Besides linear entropy, the fidelity of a state $\rho$ and the maximally mixed state $\frac{1}{d}\mathbb{I}$ is also a proper measure of mixedness.
It is defined by $M_{g}(\rho)=F(\rho,\frac{1}{d}\mathbb{I})=\frac{1}{d}(\mathrm{Tr}\sqrt{\rho})^{2}$, $0\leq M_{g}(\rho)\leq1$, and called geometric measure of mixedness.

By the definition of geometric measure of coherence $C_{g}(\rho)$, one can easily get
\begin{equation}
C_{g}(\rho)+M_{g}(\rho)\leq1.
\end{equation}
It can be verified that the equality holds for $\rho_{m}$ from (\ref{c5}) and (\ref{sqm}).
Whether there exist other form of states that satisfy the equality is still unknown, since there is no analytical formula for $C_{g}$ in general.
Thus $\rho_{m}$ is a subset of MCMS with respect to the trade-off relation between coherence measured by $C_{g}$ and mixedness measured by $M_{g}$.

For other quantifiers of coherence and mixedness, the trade-off relations remain to be investigated further. Our results may shed new light on the quantification of quantum coherence and present potential applications in quantum information theory.

\vspace{2.5ex}
\noindent{\bf Acknowledgments}\, \,
This work is supported by the National Basic Research Program of China (2015CB921002);
the National Natural Science Foundation of China Grant Nos. 11175094, 91221205 and 11275131;
Fundamental Research Funds for the Central Universities Grant No.16CX02049A; China Postdoctoral Science Foundation under Grant No.2016M600997.

\end{document}